\documentclass[preprint,prc,aps,showpacs,showkeys,groupedaddress,floatfix]{revtex4}
\usepackage{epsfig}
\usepackage{dcolumn}
\usepackage{bm}
\usepackage{graphics}
\usepackage{graphicx}
\usepackage{amssymb}
\usepackage{amsmath}
\begin{document}
\title{Any $l$-state solutions of the Hulth\'{e}n potential by the asymptotic iteration method}
\author{O. Bayrak\dag \ddag, G. Kocak\ddag \quad and I. Boztosun\ddag}
\affiliation{\dag Yozgat Faculty of Arts and Sciences, Department of
Physics, Erciyes University, Yozgat, Turkey} \affiliation{\ddag
Faculty of Arts and Sciences, Department of Physics, Erciyes
University, Kayseri, Turkey}

\date{\today}
\begin{abstract}
In this article, we present the analytical solution of the radial
Schr\"{o}dinger equation for the Hulth\'{e}n potential within the
framework of the asymptotic iteration method by using an
approximation to the centrifugal potential for any $l$ states. We
obtain the energy eigenvalues and the corresponding eigenfunctions
for different screening parameters. The wave functions are physical
and energy eigenvalues are in good agreement with the results
obtained by other methods for different $\delta $ values. In order
to demonstrate this, the results of the asymptotic iteration method
are compared with the results of the supersymmetry, the numerical
integration, the variational and the shifted $1/N$ expansion
methods.
\end{abstract}

 \maketitle

\section{Introduction}
Over the last decades, the energy eigenvalues and corresponding
eigenfunctions between interaction systems have raised a great deal
of interest in relativistic quantum mechanics as well as in
non-relativistic quantum mechanics. The exact solution of the wave
equations (relativistic or non-relativistic) are very important
since the wave function contains all the necessary information
regarding the quantum system under consideration. Analytical methods
such as the supersymmetry (SUSY) \cite{susy} and the
Nikiforov-Uvarov method (NU) \cite{Nifikorov} have been used to
solve the wave equations exactly or quasi-exactly for non zero
angular momentum quantum number ($l\neq 0$) by means of a given
potential. The radial Schr\"{o}dinger equation for the Hulth\`{e}n
potential is solved exactly by using several techniques
\cite{hulthen,nu1s,fluge} for $l=0$. For the case $l\neq0$, the
effective Hulth\'{e}n potential can not be solved exactly, but a
number of methods have been used to find the bound state energy
eigenvalues numerically \cite{Varshni} and quasi-analytically such
as the variational \cite{Varshni}, perturbation \cite{perturbation},
shifted $1/N$ expansion \cite{shift},  NU \cite{nu2s} and SUSY
\cite{susy1} methods. The Hulth\'{e}n potential \cite{hulthen} is
one of the important short-range potentials in physics and it
potential has been applied to a number of areas such as nuclear and
particle physics, atomic physics, condensed matter and chemical
physics (see \cite{Varshni} and the references therein). Therefore,
it would be interesting and important to solve the non-relativistic
radial Schr\"{o}dinger equation for this potential for $l\neq0$,
since it has been extensively used to describe the bound and the
continuum states of the interaction systems. Recently, an
alternative method, called as the asymptotic iteration method (AIM),
has been developed by \c{C}ift\c{c}i \emph{et al.}
\cite{hakanaim1,hakanaim2} for solving second-order homogeneous
linear differential equations and it has been applied to solve the
non-relativistic radial Schr\"{o}dinger equation or the relativistic
Dirac equation.

In this paper, we aim to solve the Hulth\'{e}n potential to obtain
the energy eigenvalues and corresponding eigenfunctions for any $l$
states. In the next section, AIM is introduced. Then, in section
\ref{apply1}, the Schr\"{o}dinger equation is solved by using AIM
for the Hulth\'{e}n potential for any $l$ states and our AIM results
are given in comparison with the results of the numerical
integration \cite{Varshni}, the variational \cite{Varshni}, the
shifted $1/N$ expansion \cite{shift} and the SUSY \cite{susy1}
methods. Finally, section \ref{conclude} is devoted to the summary
and conclusion.

\section{Overview of the Asymptotic Iteration Method}
\label{aim}
\subsection{Energy Eigenvalues}
AIM is briefly outlined here and the details can be found in
references \cite{hakanaim1,hakanaim2,bayrakJPA}. AIM is proposed to
solve the second-order differential equations of the form
\begin{equation}\label{diff}
  y''=\lambda_{0}(x)y'+s_{0}(x)y
\end{equation}
where $\lambda_{0}(x)\neq 0$ and the prime denotes the derivative
with respect to $x$. The variables, $s_{0}(x)$ and $\lambda_{0}(x)$,
are sufficiently differentiable. The differential equation
(\ref{diff}) has a general solution \cite{hakanaim1}
\begin{equation}\label{generalsolution}
  y(x)=exp \left( - \int^{x} \alpha(x_{1}) dx_{1}\right ) \left [C_{2}+C_{1}
  \int^{x}exp  \left( \int^{x_{1}} [\lambda_{0}(x_{2})+2\alpha(x_{2})] dx_{2} \right ) dx_{1} \right
  ]
\end{equation}
for sufficiently large $k$, $k>0$, if
\begin{equation}\label{termination}
\frac{s_{k}(x)}{\lambda_{k}(x)}=\frac{s_{k-1}(x)}{\lambda_{k-1}(x)}=\alpha(x),
\quad k=1,2,3,\ldots
\end{equation}
where
\begin{eqnarray}\label{iter}
  \lambda_{k}(x) & = &
  \lambda_{k-1}'(x)+s_{k-1}(x)+\lambda_{0}(x)\lambda_{k-1}(x) \quad
  \nonumber \\
s_{k}(x) & = & s_{k-1}'(x)+s_{0}(x)\lambda_{k-1}(x), \quad \quad
\quad \quad k=1,2,3,\ldots
\end{eqnarray}

Note that one can also start the recurrence relations from $k=0$
with the initial conditions $\lambda_{-1}=1$ and $s_{-1}=0$
\cite{fernandez}. For a given potential such as the Hulth\'{e}n, the
radial Schr\"{o}dinger equation is converted to the form of equation
(\ref{diff}). Then, s$_{0}(x)$ and $\lambda_{0}(x)$ are determined
and s$_{k}(x)$ and $\lambda_{k}(x)$ parameters are calculated by the
recurrence relations given by equation~(\ref{iter}).

The termination condition of the method in equation
(\ref{termination}) can be arranged as

\begin{equation}\label{quantization}
  \Delta_{k}(x)=\lambda_{k}(x)s_{k-1}(x)-\lambda_{k-1}(x)s_{k}(x)=0 \quad \quad
k=1,2,3,\ldots
\end{equation}

The energy eigenvalues are obtained from the roots of the equation
(\ref{quantization}) if the problem is exactly solvable. If not, for
a specific $n$ principal quantum number, we choose a suitable $x_0$
point, determined generally as the maximum value of the asymptotic
wave function or the minimum value of the potential
\cite{bayrakIJQC,hakanaim1,barakat,fernandez}, and the approximate
energy eigenvalues are obtained from the roots of this equation for
sufficiently great values of $k$ with iteration.

\subsection{Energy Eigenfunctions}
In this study, we seek the exact solution of the radial
Schr\"{o}dinger equation for which the relevant second order
homogenous linear differential equation takes the following general
form \cite{hakanaim2},
\begin{equation}
\label{compare1}
 {y}'' = 2\left( {\frac{ax^{N + 1}}{1 - bx^{N + 2}} -
\frac{\left( {t + 1} \right)}{x}} \right){y}' - \frac{w_k^t
(N)x^N}{1 - bx^{N + 2}}y, \,\,\,\,\,\,\,\,\,\,\  0<x<\infty
\end{equation}
If this equation is compared to equation (\ref{diff}), it entails
the following expressions
\begin{equation}\label{snln}
\lambda _0 (x) = 2\left( {\frac{ax^{N + 1}}{1 - bx^{N + 2}} -
\frac{\left( {t + 1} \right)}{x}} \right),   \hspace{1cm}  s_0 (x) =
- \frac{w_k^t (N)x^N}{1 - bx^{N + 2}}
\end{equation}
$a$ and $b$ are constants and $w_k^t (N)$ can be determined from the
condition (\ref{termination}) for $k=0,1,2,3,...$ and
$N$=-1,0,1,2,3,... as follows:
\begin{eqnarray}
w_k^t (-1) & = & k\left( {2a + 2bt + (k + 1)b} \right)\\
w_k^t (0) & = & 2k\left( {2a + 2bt + (2k + 1)b} \right) \\
w_k^t (1) & = & 3k\left( {2a + 2bt + (3k + 1)b} \right) \\
w_k^t (2) & = & 4k\left( {2a + 2bt + (4k + 1)b} \right) \\
w_k^t (3) & = & 5k\left( {2a + 2bt + (5k + 1)b} \right) \\
\ldots \emph{etc} \nonumber
\end{eqnarray}
Hence, these formule are easily generalized as,
\begin{equation}
w_k^t (N) = b\left( {N + 2} \right)^2k\left( {k + \frac{\left( {2t +
1} \right)b + 2a}{\left( {N + 2} \right)b}} \right)
\end{equation}
The exact eigenfunctions can be derived from the following
generator:
\begin{equation}\label{generator1}
y_n (x) = C_2 \exp \left( { - \int\limits^x
{\frac{s_{k}(x')}{\lambda_{k}(x')}dx^{'}} } \right)
\end{equation}
where $k\geq n$, $n$ represents the radial quantum number and $k$
shows the iteration number. For exactly solvable potentials, the
radial quantum number $n$ is equal to the iteration number $k$ and
the eigenfunctions are obtained directly from equation
(\ref{generator1}). For nontrivial potentials that have no exact
solutions,  $k$ is always greater than $n$ in these numerical
solutions and the approximate energy eigenvalues are obtained from
the roots of equation (\ref{quantization}) for sufficiently great
values of $k$ with iteration. It should be pointed out that
$\alpha(x)$ given by equation (\ref{termination}) is equal to zero
for the ground state. Therefore, using equation (\ref{termination})
with (\ref{snln}) in equation (\ref{generator1}), the eigenfunctions
are obtained as follows,
\begin{small}
\begin{eqnarray*}
y_0 (x) & = & C_2 \\
y_1 (x) & = & - C_2 (N + 2)\sigma \left( {1-\frac{b\left( {\rho + 1}\right)}{\sigma }x^{N + 2}} \right) \\
\\
y_2 (x) & = & C_2 (N + 2)^2\sigma \left( {\sigma + 1} \right)\left(
{1 - \frac{2b\left( {\rho + 2} \right)}{\sigma }x^{N + 2} +
\frac{b^2\left( {\rho + 2} \right)\left( {\rho + 3} \right)}{\sigma
\left( {\sigma + 1} \right)}x^{2(N + 2)}} \right)
\\
 y_3(x) & = & - C_2\frac{\sigma\left({\sigma+1}\right) \left({\sigma+2}
\right)}{\left({N+2}\right)^{-3}} \nonumber \\  & \times & \left
({1-\frac{3b\left( {\rho + 3}\right)} \sigma x^{N+2}+
\frac{3b^2\left({\rho+3}\right)\left(
{\rho+4}\right)}{\sigma\left({\sigma+1} \right)}x^{2(N+2)}
-\frac{b^3\left({\rho+3}\right)\left({\rho+4}\right)\left({\rho+ 5}
\right)}{\sigma \left( {\sigma+ 1} \right)\left( {\sigma + 2}
\right)}x^{3\left( {N + 2} \right)}} \right ) \nonumber \\
\ldots \emph{etc}
\end{eqnarray*}
\end{small}
Finally, the following general formula for the exact solutions
$y_n(x)$ is obtained as,
\begin{equation}\label{efson}
y_n (x) = \left( { - 1} \right)^nC_2 (N + 2)^n\left( \sigma
\right)_n { }_2F_1 ( - n,\rho + n;\sigma ;bx^{N + 2})
\end{equation}

It is important to note that square integrable one in $L^2$ is this
total wave function which is the asymptotic form of the wave
function times $y_n(x)$ given by equation (\ref{generator1}). Here,
$(\sigma )_n $=$\frac{\Gamma ( {\sigma + n} )}{\Gamma (\sigma) },
\quad \sigma$ = $\frac{2t + N + 3}{N + 2}$ \quad \mbox{and} \quad
$\rho$ = $\frac{( {2t + 1} )b + 2a}{( {N + 2} )b}$. The $(\sigma )_n
$ and the $_2F_1$ are known as the Pochhammer symbol and the Gauss
hypergeometric function, respectively.

\section{Calculation of the Energy Eigenvalues and Eigenfunctions}
\label{apply1} The motion of a particle with the mass $M$ in the
spherically symmetric potential is described in the spherical
coordinates by the following Schr\"{o}dinger equation:

\begin{equation}
\frac{-\hbar^{2}}{2M}\left(\frac{\partial^{2}}{\partial
r^{2}}+\frac{2}{r}\frac{\partial}{\partial r}+ \frac{1}{r^{2}}
\left[\frac{1}{\sin \theta}\frac{\partial}{\partial \theta} \left(
\sin \theta \frac{\partial}{\partial \theta} \right)
+\frac{1}{\sin^{2} \theta} \frac{\partial^{2}}{\partial \phi^{2}}
\right]+V(r) \right)\Psi_{nl m}(r,\theta,\phi)=E\Psi_{nl
m}(r,\theta,\phi)\label{sch}
\end{equation}
Defining $\Psi_{nl m}(r,\theta,\phi)=R_{nl}(r)Y_{l m}(\theta,\phi)$,
we obtain the radial part of the Schr\"{o}dinger equation:
\begin{eqnarray}
\left(\frac{d^{2}}{d{r}^{2}}+\frac {2}{r}\frac{d}{dr}\right)R_{nl}(r)
+\frac{2M}{\hbar^{2}}\left[E-V(r)-\frac{l(l+1)\hbar^{2}}{2 M r^{2}} \right]R_{nl}(r)=0
\label{radialr}
\end{eqnarray}
It is sometimes convenient to define $R_{nl}(r)$ and the
effective potential as follows
\begin{equation}
R_{nl}(r)=\frac{u_{nl}(r)}{r}, \quad
V_{eff}=V(r)+\frac{l(l+1)\hbar^{2}}{2 M r^{2}}
\end{equation}
Since
\begin{equation}
\left(\frac{d^{2}}{d{r}^{2}}+\frac {2}{r}\frac{d}{dr}
\right)\frac{u_{nl}(r)}{r}=\frac{1}{r}\frac{d^{2}}{d{r}^{2}}u_{nl}(r)
\end{equation}
The radial Schr\"{o}dinger equation \cite{fluge} given by equation
(\ref{radialr}) follows that
\begin{equation}
\frac{d^{2}u_{nl}(r)}{d{r}^{2}}+\frac {2M}{\hbar^{2}}
\left[E-V_{eff} \right]u_{nl}(r)=0 \label{factorschrodinger}
\end{equation}

\begin{figure}[h]
\includegraphics{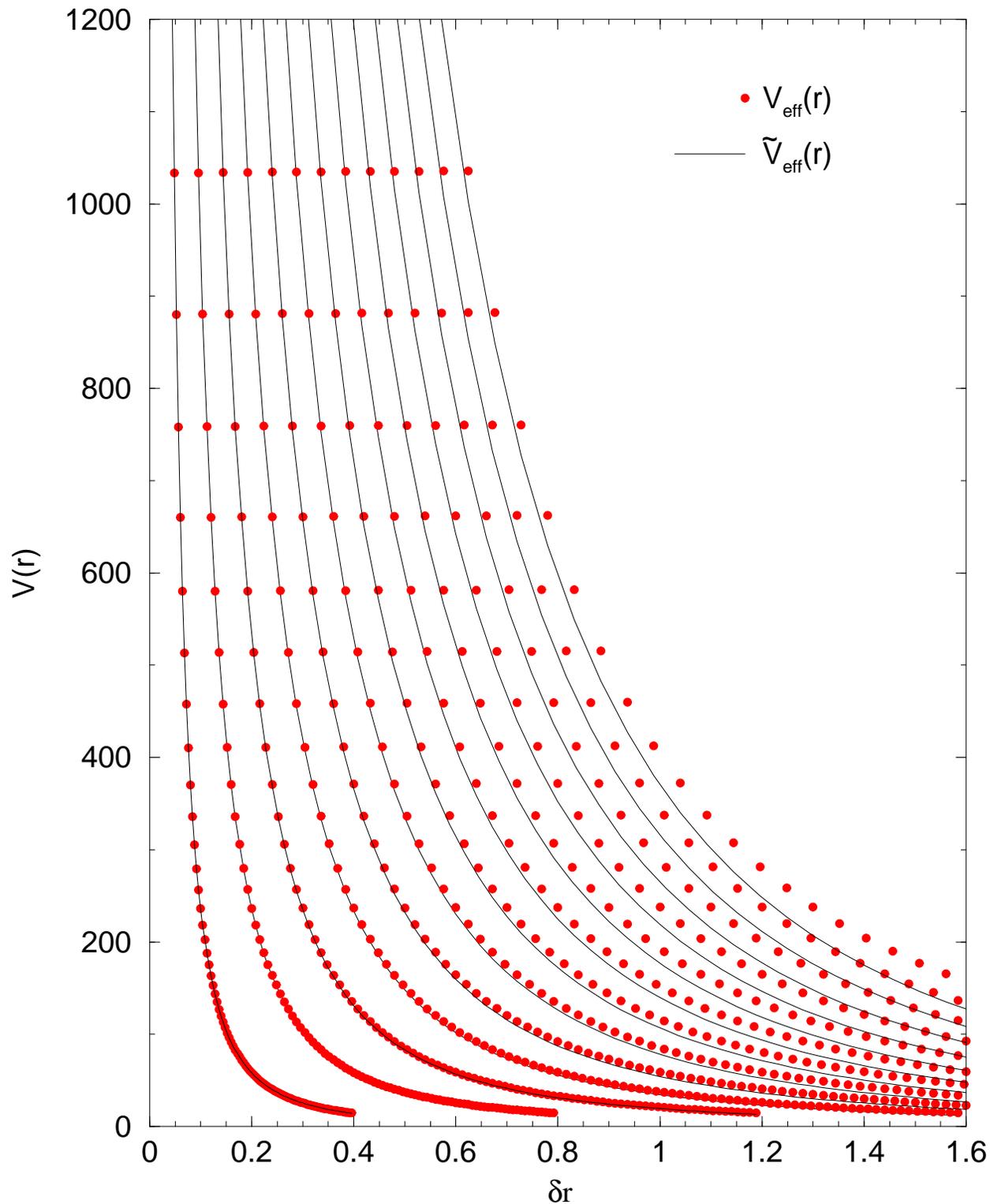}
\caption{The variation of the effective Hulth\'{e}n ${V}_{eff}(r)$
 and the approximate effective Hulth\'{e}n
potentials $\widetilde{V}_{eff}(r)$ with respect to the $\delta r$
for various values of the screening parameter. The parameters are in
atomic units $(\hbar=e=m=1)$ and $\delta$ change from 0.4 to 5.2 in
steps of 0.4.} \label{eigenpot}
\end{figure}

The Hulth\'{e}n potential \cite{hulthen} is given by

\begin{equation}\label{dh}
  V_{H}(r)=-Ze^{2}\delta\frac{e^{-\delta r}}{1-e^{-\delta r}}
\end{equation}
where Z and $\delta$ are respectively the atomic number and the
screening parameter, determining the range for the Hulth\'{e}n
potential. The Hulth\'{e}n potential behaves like the Coulomb
potential near the origin $(r\longrightarrow0)$, but in the
asymptotic region $(r\gg1)$, the Hulth\'{e}n potential decreases
exponentially, so its capacity for bound states is smaller than the
Coulomb potential. However, for small values of the screening
parameter $\delta$, the Hulth\'{e}n potential becomes the Coulomb
potential given by $V_{C}=-\frac{Ze^{2}}{r}$. The effective
Hulth\'{e}n potential is

\begin{equation}\label{veff}
    V_{eff}(r)=V_{H}(r)+V_{l}=-Ze^{2}\delta \frac{e^{-\delta r}}{1-e^{-\delta
    r}}+\frac{l(l+1)\hbar^{2}}{2Mr^{2}}
\end{equation}
where $V_{l}=\frac{l(l+1)\hbar^{2}}{2Mr^{2}}$ is known as the
centrifugal term. This effective potential can not be solved
analytically for $l\neq0$ because of the centrifugal term.
Therefore, we must use an approximation for the centrifugal term
similar to other authors
\cite{susy1,nu2s,other1,other2,other3,other4}. In this
approximation, $\frac{1}{r^{2}}=\delta^{2}\frac{e^{-\delta
r}}{(1-e^{-\delta r})^{2}}$ is used for the centrifugal term. As
shown in figure \ref{eigenpot}, this approximation is only valid for
small $\delta r$ and it breaks down in the high-screening region.
For small $\delta r$, $\widetilde{V}_{eff}(r)$ is very well
approximated to $V_{eff}(r)$ and the Schr\"{o}dinger equation for
this approximate potential is solvable analytically. So, the
effective potential becomes

\begin{equation}\label{veff1}
    \widetilde{V}_{eff}(r)=-Ze^{2}\delta \frac{e^{-\delta r}}{1-e^{-\delta
    r}}+\frac{l(l+1)\hbar^{2}\delta^{2}}{2M}\frac{e^{-\delta r}}{(1-e^{-\delta
    r})^{2}}
\end{equation}

Instead of solving the radial Schr\"{o}dinger equation for the
effective Hulth\'{e}n potential $V_{eff}(r)$ given by equation
(\ref{veff}), we now solve the radial Schr\"{o}dinger equation for
the new effective potential $\widetilde{V}_{eff}(r)$ given by
equation (\ref{veff1}). Inserting this new effective potential into
equation (\ref{factorschrodinger}) and using the following
\emph{ansatzs} in order to make the differential equation more
compact
\begin{equation}\label{ansatz}
-\varepsilon^{2}=\frac{2M E}{\hbar^{2}\delta^{2}}, \quad
\beta^{2}=\frac{2M Z e^{2}}{\hbar^{2}\delta}, \quad \delta r=x
\end{equation}
The radial Schr\"{o}dinger equation takes the following form:
\begin{equation}\label{radyal}
 \frac{d^{2}u_{nl}(x)}{dx^{2}}+
 \left[ -\varepsilon^{2}+\beta^{2}\frac{e^{-x}}{(1-e^{-x})}-l(l+1)\frac{e^{-x}}{(1-e^{-x})^{2}} \right] u_{nl}(x)=0
\end{equation}
If we rewrite equation (\ref{radyal}) by using a new variable of the
form $z=e^{-x}$, we obtain
\begin{equation}\label{trans}
\frac{d^{2}u_{nl}(z)}{dz^{2}}+\frac{1}{z}\frac{du_{nl}(z)}{dz}
+\left[-\frac{\varepsilon^{2}}{z^{2}}+
\frac{\beta^{2}}{z(1-z)}-\frac{l(l+1)}{z(1-z)^{2}} \right]u_{nl}(z)=0
\end{equation}
In order to solve this equation with AIM, we should
transform this equation to the form of equation (\ref{diff}).
Therefore, the reasonable physical wave function we propose is as
follows
\begin{equation}\label{wave}
u_{nl}(z)=z^{\varepsilon} (1-z)^{l+1}f_{nl}(z)
\end{equation}
If we insert this wave function into equation (\ref{trans}), we have
the second-order homogeneous linear differential equations as in the
following form
\begin{equation}\label{aimschr}
\frac{d^{2}f_{nl}(z)}{dz^{2}}=\left[\frac{(2\varepsilon+2l+3)z-(2\varepsilon+1)}{z(1-z)}\right]
\frac{df_{nl}(z)}{dz}+\left[\frac{(2\varepsilon+l+2)l+2\varepsilon-\beta^{2}+1}{z(1-z)}
\right]f_{nl}(z)
\end{equation}
which is now amenable to an AIM solution. By comparing this equation
with equation (\ref{diff}), we can write the $\lambda_{0}(z)$ and
$s_{0}(z)$ values and by means of equation (\ref{iter}), we may
calculate $\lambda_k(z)$ and $s_k(z)$. This gives:
\begin{eqnarray}\label{ini}
    \lambda_{0}(z)&=&\left(\frac{(2\varepsilon+2l+3)z-(2\varepsilon+1)}{z(1-z)}\right) \nonumber \\
    s_{0}(z)&=&\left(\frac{(2\varepsilon+l+2)l+2\varepsilon-\beta^{2}+1}{z(1-z)} \right)\nonumber \\
    \lambda_{1}(z)&=&\frac{2+6\,\varepsilon-7\,z-2\,lz-{\beta}^{2}z+12\,{z}^{2}l-18\,
\varepsilon\,z-6\,\varepsilon\,zl}{{z}^{2} \left( -1+z \right) ^{2}} \nonumber \\&+ &
\frac{12\,\varepsilon\,{z}^{2}+11\,{z}^{2}+4\,{\varepsilon}^{2}+{l}^{2}z+{\beta}^{2}
{z}^{2}+4\,{\varepsilon}^{2}{z}^{2}-8\,{\varepsilon}^{2}z+6\,\varepsilon\,{z}^{2}l+3\,{l}^{2}{z}^{2}}{{z}^{2}
 \left( -1+z \right) ^{2}}\nonumber \\
    s_{1}(z)&=&{\frac { \left(2\,l+2\,\varepsilon-{\beta}^{2}+2\,\varepsilon\,l+{l}^{2}+1\right)
    \left( -2+5\,z+2\,\varepsilon\,z+2\,lz-2\,\varepsilon \right) }{{z}^{2} \left( -1+z \right) ^{2}}}\nonumber \\
     \ldots \emph{etc}
\end{eqnarray}
Combining these results with the quantization condition given by
equation (\ref{quantization}) yields
\begin{eqnarray}
 s_{0}\lambda_{1}-s_{1 }\lambda_{0}&=0& \,\,\,\,\,\, \Rightarrow
\,\,\,\,\,\,\varepsilon_{0}  =  \frac{\beta^{2}-1-2l-l^{2}}{2(l+1)},  \quad for \quad k=1   \nonumber \\
 s_{1}\lambda_{2}-s_{2}\lambda_{1}&=0& \,\,\,\,\,\, \Rightarrow
\,\,\,\,\,\, \varepsilon_{1}= \frac{\beta^{2}-4-4l-l^{2}}{2(l+2)}, \quad for \quad k=2  \nonumber \\
 s_{2}\lambda_{3}-s_{3}\lambda_{2}&=0& \,\,\,\,\,\, \Rightarrow
\,\,\,\,\,\,\varepsilon_{2}=\frac{\beta^{2}-9-6l-l^{2}}{2(l+3)}, \quad for \quad k=3  \\
\ldots \emph{etc} \nonumber
 \end{eqnarray}
When the above expressions are generalized, the eigenvalues turn out
as
\begin{equation}\label{energy}
\varepsilon_{nl}=\left( \frac{\beta^{2}-(n+l+1)^{2}}{2(n+l+1)}\right)
\hspace{1cm} n,l=0,1,2,3,...
\end{equation}
Using equation (\ref{ansatz}), we obtain the energy eigenvalues
E$_{nl}$,
\begin{equation}\label{energyeigenvalues}
E_{nl}=-\frac{\hbar^{2}}{2M}\left[\frac{MZe^{2}}{\hbar^{2}(n+l+1)}-\frac{(n+l+1)\delta}{2}\right]^{2}
\hspace{1cm}
\end{equation}
In the atomic units $(\hbar =M =e = 1)$ and for Z = 1, equation
(\ref{energyeigenvalues}) turns out to be
\begin{equation}\label{eigenvaluesatomic}
E_{nl}=-\frac{1}{2}\left[\frac{1}{(n+l+1)}-\frac{(n+l+1)\delta}{2}\right]^{2}
\hspace{1cm}
\end{equation}

In order to test the accuracy of equation (\ref{eigenvaluesatomic}),
we calculate the energy eigenvalues for $Z=1$, any $n$ and $l$
quantum numbers and several values of the screening parameter. AIM
results are compared with the results of the numerical integration
\cite{Varshni}, the variational \cite{Varshni}, the shifted $1/N$
expansion \cite{shift} and the SUSY \cite{susy1} methods in Table
\ref{Table1} and Table \ref{Table2}. As it can be seen from the
results presented in these tables, the AIM results are in good
agrement with the results of the other methods for the small
$\delta$ values. For large $\delta$ values, there are differences
between our results and the results of others. This difference is
due to the $\widetilde{V}_{eff}(r)$ potential, which we have used to
approximate the $V_{eff}(r)$ potential. As it is seen from figure
\ref{eigenpot}, for large $\delta r$ values, the discrepancy becomes
apparent between our $\widetilde{V}_{eff}(r)$ and the true
$V_{eff}(r)$ potentials. This gives rise to the differences for the
eigenvalues presented in Tables \ref{Table1} and \ref{Table2} at
large $\delta$ values.

Now, as indicated in Section \ref{aim}, we can determine the
corresponding wave functions by using equation (\ref{efson}). When
we compare equation (\ref{compare1}) and equation (\ref{aimschr}),
we find $N=-1$, $b=1$, $a=l+1$, and $t=\frac{2\varepsilon-1}{2}$.
Therefore, we find $\rho=2(\varepsilon+l+1)$ and
$\sigma=2\varepsilon+1$. So, we can easily find the solution for
$f_{nl}(z)$, for the energy eigenvalue equation
(\ref{energyeigenvalues}) by using equation (\ref{efson}).
\begin{equation}
f_{nl}(z)=(-1)^{n}\frac{\Gamma(2\varepsilon_{n}+n+1)}{\Gamma(2\varepsilon_{n}+1)}
{_{2}}F_{1}(-n,2\varepsilon_{n}+2l+2+n;2\varepsilon_{n}+1;z)
\end{equation}
Thus, we can write the total radial wave function as below,
\begin{equation}\label{radyalwave}
u_{nl}(z)=Nz^{\varepsilon_{n}}(1-z)^{l+1}{_{2}}F_{1}(-n,2(\varepsilon_{n}+l+1)+n;2\varepsilon_{n}+1;z)
\end{equation}
where $N$ is the normalization constant.

\section{Conclusion}
\label{conclude} We have shown an alternative method to obtain the
energy eigenvalues and corresponding eigenfunctions of the
Hulth\'{e}n potential within the framework of the asymptotic
iteration method for any $l$ states. We have calculated the energy
eigenvalues for the Hulth\'{e}n potential with $Z=1$ and several
values of the screening parameter. The wave functions are physical
and energy eigenvalues are in good agreement with the results
obtained by other methods. In order to demonstrate this, AIM results
have been compared with the results of the numerical integration
\cite{Varshni}, the variational \cite{Varshni}, the shifted $1/N$
expansion \cite{shift} and the SUSY \cite{susy1} methods in Tables
\ref{Table1} and \ref{Table2}. For small $\delta$ values, AIM
results are in good agreement with the results of the other methods,
but in the high screening region, the agreement is poor. The reason
is simply that when the $\delta r$ increases in the high screening
region, the agreement between $V_{eff}(r)$ and
$\widetilde{V}_{eff}(r)$ potentials decreases as shown in Figure
\ref{eigenpot}. This problem could be solved by making a better
approximation of the centrifugal term.

It should be pointed out that the asymptotic iteration method gives
the eigenvalues directly by transforming the radial Schr\"{o}dinger
equation  into a form of ${y}''$ =$ \lambda _0 (r){y}' + s_0 (r)y$.
The wave functions are easily constructed by iterating the values of
$s_0(r)$ and $\lambda_0(r)$. The asymptotic iteration method results
in exact analytical solutions if there is and provides the
closed-forms for the energy eigenvalues as well as the corresponding
eigenfunctions. Where there is no such a solution, the energy
eigenvalues are obtained by using an iterative approach
\cite{bayrakIJQC,barakat,fernandez}. As it is presented, AIM puts no
constraint on the potential parameter values involved and it is easy
to implement. The results are sufficiently accurate for practical
purposes.

\section*{Acknowledgments} This paper is an output of the project supported by the
Scientific and Technical Research Council of Turkey
(T\"{U}B\.{I}TAK), under the project number TBAG-106T024 and Erciyes
University (FBA-03-27, FBT-04-15, FBT-04-16). Authors would also
like to thank Dr. Ay\d{s}e Boztosun for the proofreading as well as
Professor R. L. Hall and Dr. H. \c{C}ift\c{c}i for useful comments
on the manuscript.

\newpage
\begin{table}[tbp]
\begin{center}
\begin{tabular}{ccccccccccccccc}
\hline\hline$State$&  & $\delta$& &  AIM& &SUSY \cite{susy1} & &
Numerical Int. \cite{Varshni}& & Variational \cite{Varshni}& &
Shifted $\frac{1}{N}$ \cite{shift}
\\\hline 2p
  & &0.025& &0.1128125& &0.1127605& &0.1127605& & 0.1127605& &               \\
  & &0.050& &0.1012500& &0.1010425& &0.1010425& & 0.1010425& &0.1010424       \\
  & &0.075& &0.0903125& &0.0898478& &0.0898478& & 0.0898478& &               \\
  & &0.100& &0.0800000& &0.0791794& &0.0791794& & 0.0791794& &0.0791794       \\
  & &0.150& &0.0612500& &0.0594415& &0.0594415& & 0.0594415& &               \\
  & &0.200& &0.0450000& &0.0418854& &0.0418860& & 0.0418860& &0.0418857       \\
  & &0.250& &0.0312500& &0.0266060& &0.0266111& & 0.0266108& &               \\
  & &0.300& &0.0200000& &0.0137596& &0.0137900& & 0.0137878& &               \\
  & &0.350& &0.0112500& &0.0036146& &0.0037931& & 0.0037734& &               \\\\
3p
  & &0.025& &0.0437590& &0.0437068& &0.0437069& & 0.0437069& &                \\
  & &0.050& &0.0333681& &0.0331632& &0.0331645& & 0.0331645& &0.03316518       \\
  & &0.075& &0.0243837& &0.0239331& &0.0239397& & 0.0239397& &                \\
  & &0.100& &0.0168056& &0.0160326& &0.0160537& & 0.0160537& &0.01606772       \\
  & &0.150& &0.0058681& &0.0043599& &0.0044663& & 0.0044660& &               \\\\
3d
  & &0.025& &0.0437587& &0.0436030& &0.0436030& & 0.0436030& &                \\
  & &0.050& &0.0333681& &0.0327532& &0.0327532& & 0.0327532& &0.0327532        \\
  & &0.075& &0.0243837& &0.0230306& &0.0230307& & 0.0230307& &                \\
  & &0.100& &0.0168055& &0.0144832& &0.0144842& & 0.0144842& &0.0144842        \\
  & &0.150& &0.0058681& &0.0132820& &0.0013966& & 0.0013894& &                \\\\\hline\hline
\end{tabular}
\end{center}
\caption{Energy eigenvalues of the Hulth\'{e}n potential as a
function of the screening parameter for 2p, 3p and 3d states in
atomic units $(\hbar =m =e =1)$ and for $Z=1$.} \label{Table1}
\end{table}

\begin{table}[tbp]
\begin{center}
\begin{tabular}{ccccccccccccccc}
\hline\hline$State$&  & $\delta$& &  AIM &  &  SUSY \cite{susy1} & &
Numerical Int. \cite{Varshni}& & Variational \cite{Varshni}& &
Shifted $\frac{1}{N}$ \cite{shift}
\\\hline
 4p
  & &0.025& &0.0200000& &0.0199480& &0.0199489& & 0.0199489& &                \\
  & &0.050& &0.0112500& &0.0110430& &0.0110582& & 0.0110582& &0.0110725        \\
  & &0.075& &0.0050000& &0.0045385& &0.0046219& & 0.0046219& &                \\
  & &0.100& &0.0012500& &0.0004434& &0.0007550& & 0.0007532& &                \\
4d
  & &0.025& &0.0200000& &0.0198460& &0.0198462& & 0.0198462& &                \\
  & &0.050& &0.0112500& &0.0106609& &0.0106674& & 0.0106674& &0.0106690        \\
  & &0.075& &0.0050000& &0.0037916& &0.0038345& & 0.0038344& &                \\
4f
  & &0.025& &0.0200000& &0.0196911& &0.0196911& & 0.0196911& &                \\
  & &0.050& &0.0112500& &0.0100618& &0.0100620& & 0.0100620& &0.0100620        \\
  & &0.075& &0.0050000& &0.0025468& &0.0025563& & 0.0025557& &                \\\\
5p
  & &0.025& &0.0094531& &0.0094011& &0.0094036& &         & &0.0094087        \\
  & &0.050& &0.0028125& &0.0026056& &0.0026490& &         & &                \\
5d
  & &0.025& &0.0094531& &0.0092977& &0.0093037& &         & &0.0093050        \\
  & &0.050& &0.0028125& &0.0022044& &0.0023131& &         & &                \\
5f
  & &0.025& &0.0094531& &0.0091507& &0.0091521& &         & &0.0091523        \\
  & &0.050& &0.0028125& &0.0017421& &0.0017835& &         & &                \\
5g
  & &0.025& &0.0094531& &0.0089465& &0.0089465& &         & &0.0089465        \\
  & &0.050& &0.0028125& &0.0010664& &0.0010159& &         & &                \\\\
6p
  & &0.025& &0.0042014& &0.0041493& &0.0041548& &         & &                \\
6d
  & &0.025& &0.0042014& &0.0040452& &0.0040606& &         & &                \\
6f
  & &0.025& &0.0042014& &0.0038901& &0.0039168& &         & &               \\
6g
  & &0.025& &0.0042014& &0.0036943& &0.0037201& &         & &               \\\hline\hline
\end{tabular}
\end{center}
\caption{Energy eigenvalues of the Hulth\'{e}n potential as a
function of the screening parameter for 4p, 4d, 4f, 5p, 5d, 5f ,5g,
6p, 6d, 6f and 6g states in atomic units $(\hbar =m =e =1)$ and for
$Z=1$.} \label{Table2}
\end{table}

\end{document}